\documentclass[11pt]{article}
\usepackage{amssymb}
\usepackage{txfonts}%
\usepackage{graphicx}
\usepackage{epsfig}
\usepackage{graphics,color}
\usepackage{amsbsy}

\newcommand{\be}{\begin{equation}}
 \newcommand{\ee}{\end{equation}}
 \newcommand{\bse}{\begin{subequations}}
 \newcommand{\ese}{\end{subequations}}
 \newcommand{\bea}{\begin{eqnarray}}
 \newcommand{\eea}{\end{eqnarray}}

\newcommand{\bean}{\begin{eqnarray*}}
\newcommand{\eean}{\end{eqnarray*}}

\begin{document}

\begin{center}

{\bf Multisource thermal model to the transverse momentum spectra in
$pp$ collisions at RHIC and LHC energies} \vskip1.0cm

Bao-Chun Li$^{1, 2}$\footnote{Corresponding Author, E-mail:
libc2010@163.com}, Hai-Fu Zhao$^{1}$, Fu-Hu Liu$^{1, 2}$, Xin-Jian
Wen$^{1,2}$ and Hong-Wei Dong$^{1,2}$

{\small {\it $^{1}$College of Physics and Electronics Engineering,
State Key Laboratory of Quantum Optics and Quantum Optics Devices,
Shanxi University, Taiyuan 030006, China
\\$^{2}$Collaborative
Innovation Center of Extreme Optics, Shanxi University, Taiyuan
030006, China }}

\end{center}

\begin{abstract}

In this paper, an improved multi-source thermal model is used to
analyze the transverse momentum spectra in $pp$ collisions at high
energies ranging from $\sqrt{\mathrm{\it s_{NN}}}$ = 62.4 GeV to 7
TeV. We give a detailed comparison between the theoretical results
and experimental data at RHIC and LHC energies. It is shown that the
excitation factors of emission sources depend linearly on
ln$\sqrt{\mathrm{\it s_{NN}}}$ in the framework. Based on the
variation regularity of the source-excitation factors, transverse
momentum spectra are predicted in $pp$ collisions at higher
energies, potential future $pp$ colliders operating at
$\sqrt{\mathrm{\it s_{NN}}}$ = 33 and 100 TeV.\\
PACS number(s): 13.85.-t, 14.40.-n, 12.40.Ee

\end{abstract}

\vskip1.0cm

\newpage

{\section{Introduction}}

  Inclusive measurements of charged-particle spectra
in $pp$ collisions can provide an insight into the strong
interaction in low-energy, non-perturbative region of Quantum
Chromodynamics (QCD)~\cite{Aad:2016mok}. Investigations of the
charged-particle production are refining the understanding of global
properties of $pp$ collisions at the LHC. As the collision energy
increases, a much broader and deeper study of QGP will be done at
the LHC. It leads to a significant extension of the kinematic range
in longitudinal rapidity and transverse momentum. And, the
charged-particle distribution is very helpful in understanding the
basic production mechanism of hadrons produced in nucleon-nucleon
and nucleus-nucleus collision experiments. Transverse momentum
spectra of the particles play an important role in the observation
of high-energy collisions. The spectra can provide insight into
particle production as well as matter evolution in $pp$ or $AA$
collisions at RHIC and LHC energies~\cite{Mishra:2015pta}.

At one of the later stages in collisions, the system will be
dominated by hadronic resonances.  At the last stage of high-energy
collisions, the interacting system at the kinetic freeze-out stays
at a thermodynamic equilibrium state or local equilibrium state. The
particle emission process is influenced by not only the thermal
motion but also the collective flow.  Considering the creation and
subsequent decay of hadronic resonances produced in chemical
equilibrium at unique temperature and baryon chemical potential,
thermal-statistical models have given a consistent description of
particle production in heavy-ion collisions at high energies during
the past two decades. The identifying feature of the thermal model
is that all the resonances as listed by the Particle Data Group are
assumed to be in thermal and chemical
equilibrium~\cite{Cleymans:2016qnc}. Statistical thermal models have
successfully described the particle abundances at low
$P_T=\sqrt{P_x^2+P_y^2}$~\cite{Becattini:2000jw, Broniowski:2002ea,
Broniowski:2002nf}.

In recent years, some phenomenological models or (semi-) empirical
formulas of particle distributions have been reported to explain the
experimental data of the $P_T$ (or $m_T=\sqrt{P_T^2+m^2}$) spectra
in $pp$ and $AA$ collisions, up to LHC energies. It is interesting
to note that various exponential functions in the distribution
descriptions were adopted~\cite{Abelev:2008ab}, such as $P_T$
exponential distribution $\frac{dN}{P_TdP_T}=c\exp(-P_T/T_{P_T})$,
$P_T^2$ exponential or $P_T$ Gaussian
$\frac{dN}{P_TdP_T}=c\exp(-P_T^2/T_{P_T}^2)$ and $P_T^3$ exponential
distribution $\frac{dN}{P_TdP_T}=c\exp(-P_T^3/T_{P_T}^3)$, $m_T$
exponential distribution $\frac{dN}{m_Tdm_T}=c\exp(-m_T/T_{m_T})$,
Boltzmann distribution $\frac{dN}{m_Tdm_T}=cm_T\exp(-m_T/T_{B})$ and
Bose-Einstein distribution
$\frac{dN}{m_Tdm_T}=c/[\exp(m_T/T_{BE})-1]$.

In this work, combined with the exponential functions, we improve a
multi-source thermal model by considering the rapidity shifts of
longitudinal sources along the rapidity axis. The improved model is
used to analyze the transverse momentum spectra in $pp$ collisions
at RHIC and LHC energies.

{\section{The description model}}

 In order to understand the particle
spectra observed in multiparticle production processes, Hagedorn
proposed a statistical description~\cite{Hagedorn:1965st}, where the
transverse momentum $P_T$ spectra follow a thermalized Boltzmann
type of distribution

 \bea f(P_T)=\frac{dN}{P_TdP_T}=c
\exp\left[-\frac{P_{T}}{<P_{T}>}\right]\,\,,\label{exponential}
 \eea
where the $<\cdot \cdot \cdot >$ denotes the
 mean transverse momentum averaged over all events in the event sample.
 It is an exponential function and can only fit the experimental
data in a limited range of transverse mass, $0.2< m_T-m_\pi < 0.7$
GeV~\cite{Abgrall:2013qoa}. But, a width of the rapidity (or
pseudorapidity) distribution of corresponding particles is not
considered for the $P_T$ distribution function. In order to be
consistent with experimental data, the pseudorapidity interval
integral has been added,
\begin{equation}
\left(E\frac{d^3N}{dP^3}\right)_{\eta}=\int_{\eta_{min}}^{\eta_{max}}d\eta
\frac{dy}{d\eta}\left(E\frac{d^3N}{dP^3}\right)\,\,,
\end{equation}
where
\begin{equation}
\frac{dy}{d\eta}(\eta, P_T)=\sqrt{1-\frac{m^2}{m^2_T\cosh^2 y}}\,\,.
\end{equation}
The rapidity $y$ is a function of $\eta$ and $P_T$
\begin{equation}
y=\frac{1}{2}\ln \Big [\frac{\sqrt{P_T^2 \cosh^2 \eta +
m^2}+P_T\sinh \eta}{\sqrt{P_T^2 \cosh^2 \eta + m^2}-P_T\sinh
\eta}\Big]\,\,.
\end{equation}

Then, we resolve the issue by a multisource thermal model. According
to the geometrical picture of high-energy collisions, the
thermalized cylinder model~\cite{Liu:2002ws} and the relativistic
diffusion model~\cite{Wolschin:2011mz}, particle emission sources
located in the projectile and target cylinders are formed in $pp$
collisions. At intermediate energy, the two cylinders overlap
totally and are regarded as a single cylinder. At high energy, the
two cylinders overlap partly. At ultra-high energy, the two
cylinders are completely separate, resulting in a gap between them.
In the rapidity $y$ space, the projectile cylinder and the target
cylinder lie in the rapidity ranges $[y_{Pmin}, y_{Pmax} ]$ and
$[y_{Tmin}, y_{Tmax}]$, respectively. The center rapidity of the
interacting system is denoted by $y_C$ (or $y_{cm}$). So, the
rapidity distribution of particles produced in the collision is
given by
 \bea f(y)
=\frac{k_t}{y_{tmax}-y_{tmin}}\int^{y_{tmax}}_{y_{tmin}}f_{s}(y,
y')d{y'} + \frac{k_p}{y_{pmax}-y_{pmin}}\int^{y_{pmax}}_{y_{pmin}}
f_{s}(y, y')d{y'}\,\,, \label{c}
 \eea
where $f_{s}(y, y')$ is the rapidity distribution of particles
emitted from a source at $y'$. In terms of the description of the
transverse momentum $P_T$ and the transverse mass $m_T$
distributions, we need to consider the longitudinal rapidity of the
emission source in the cylinder(s). These sources with different
rapidity shifts in the rapidity space located nonuniformly in the
rapidity region. In order to deal conveniently with the relation
between the sources and particles, the sources can be divided into
$n$ groups in accordance with the longitudinal locations. Due to
different interaction mechanisms or event samples, the source number
in the $i$th group is assumed to be $k_i$ . Identified fragments or
particles emit isotropically from different emission points (also
known as sources) formed in the high-energy collisions. The
transverse momentum spectrum contributed by the $j$th source in the
$i$th group is an exponential distribution
 \bea f_{ij}(P_{Tij})=
\frac{1}{<P_{Tij}>}\exp\left[-\frac{P_{Tij}}{<P_{Tij}>}\right],\label{b}
 \eea
where a source-excitation factor
 \bea <P_{Tij}>=\int
P_{Tij}f_{ij}(P_{Tij})dP_{Tij} \label{b} \eea
 is the
mean value of the transverse momentum of particles which come from
the given source in the  group. In the same group, we have
   \bea  <P_{Ti1}> = <P_{Ti2}> =\cdot\cdot\cdot  =<P_{Tik_i}> =<P_{Ti}>. \eea
  By computing the convolution of the exponential function Eq. (1),
  the transverse momentum distribution contributed by the $i$th group is
   \bea f_i(P_T)=
\frac{P_T^{k_i-1}}{(k_i-1)!<P_{Ti}>^{k_i}}\exp\left[-\frac{P_{T}}{<P_{Ti}>}\right].\label{b}
 \eea
It is an Erlang distribution. The transverse momentum distribution
is
 \bea f(P_T)=\sum_{i=1}^nc_i f_i(P_T)\,\,, \label{b}
 \eea
which is known as a multi-component Erlang distribution. The $c_i$
is the share of the $i$th group. In the improved model, the rapidity
cut is naturally and consistently taken into account. To simplify
the calculation, the Monte Carlo method is used to obtain the
transverse momentum spectrum. With Eq. (1), the transverse momentum
is
  \bea  P_{Tij}=-<P_{Tij}>\ln R_{ij}\,\,, \eea
where $R_{ij}$ is a random number in [0, 1].

{\section{Results and Discussion}}

In order to identify the validity of the distribution function, Eq.
(5), figure 2 give pseudorapidity distributions of charged particles
produced in $pp$ collisions at $\sqrt{\mathrm{\it s_{NN}}}$ =200
GeV. Filled circles represent experimental data measured by the
PHOBOS Collaboration~\cite{Alver:2010ck}. The theoretical result is
presented by a curve. The $\chi^2$ per number of degrees of freedom
($\chi^2$/ndf) testing provides statistical indication of the most
probable value of corresponding parameters. We see that the
calculated results agree with the experimental data.

Figure 2 shows the transverse momentum spectra of identified charged
hadrons (pion, kaon, proton) in $pp$ collisions at
$\sqrt{\mathrm{\it s_{NN}}}$ = 62.4 GeV and 200 GeV. The scattered
symbols present the experimental data from the PHENIX
Collaboration~\cite{Adare:2011vy, Adare:2008qb, Adare:2010fe} and
the STAR Collaboration~\cite{Abelev:2006cs, Adams:2006nd}. The solid
lines present the model results. Our results for $P_T$ spectra are
in good agreement with the experimental data. The maximum
$\chi^2$/ndf value is 1.15 and the minimum $\chi^2$/ndf value is
0.08. Figure 3 shows the transverse momentum spectra of identified
charged hadrons (pion, kaon, proton) in $pp$ collisions in the range
$|y|<1$, at $\sqrt{\mathrm{\it s_{NN}}}$ = 0.9, 2.76 and 7 TeV. The
scattered symbols present the experimental data from the CMS
Collaboration~\cite{Chatrchyan:2012qb}. The solid lines present the
model results, which are in good agreement with the experimental
data. The maximum value of $\chi^2$/ndf is 1.04 and the minimum
value is 0.10.

 According to the pseudorapidity distributions, the sources may be divided into two groups $n=2$. The
 parameter values are obtained by fitting the experimental
data. In the calculations, we take one source in the first group
$k_1=1$ and two sources in the second group $k_2=2$. For pions,
kaons and protons, the mean $P_T$ in the second group $<P_{T2}>$ are
fixed, i.e., 0.11 GeV/c, 0.20 GeV/c and 0.26 GeV/c, respectively. As
can be seen in Fig. 4, the mean $P_T$ of the first group $<P_{T1}>$
varies regularly with ln$\sqrt{\mathrm{\it s_{NN}}}$,
$<P_{T1}>=(0.0246\pm0.006)$ln$\sqrt{\mathrm{\it
s_{NN}}}$$+(0.180\pm0.011)$,
$<P_{T1}>=(0.0728\pm0.002)$ln$\sqrt{\mathrm{\it
s_{NN}}}$$+(0.019\pm0.001)$ and
$<P_{T1}>=(0.0755\pm0.0003)$ln$\sqrt{\mathrm{\it
s_{NN}}}$$+(0.020\pm0.002)$ for pions, kaons and protons,
respectively. Based on the linear functions, we can predict the
$<P_{T1}>$ taken in the model for $pp$ collisions at higher
energies.  When $\sqrt{\mathrm{\it s_{NN}}}$ is increased to 33 and
100 TeV, the $<P_{T1}>$ values for pions, kaons and protons are
taken to be 0.4354 and 0.4626, 0.7766 and 0.8574, 0.8051 and 0.8887,
respectively. The prediction of the transverse momentum spectra of
pions, kaons and protons are given in Fig. 5.

{\section{Conclusions}}

Final-state particles produced in high energy collisions have
attracted much attention, since attempt have been made to understand
the properties of strongly coupled QGP by studying the possible
production mechanisms. Many exponential distributions are suggested
in description of the final-state particle distribution. Thermal
statistical models have been successful in describing particle
production in various systems at different
energies~\cite{Tiwari:2013wga, Huang:2003jv}. In our previous
work~\cite{Liu:2014xna}, to extract the chemical potentials of
quarks from ratios of negatively/positively charged particles, we
have analyzed the transverse momentum spectrums of the CMS and ALICE
Collaborations by using the Tsallis distribution. The results in low
transverse momentum region in the present work is better than those
in our previous work. Based on different formula on transverse
momentum spectrum in the model, the dependences of transverse
momentum, centrality and participant nucleon number  on elliptic
flow in GeV and TeV energy regions have been studied in our previous
works~\cite{Wang:2011zzh, Li2013AHEP}. Comparing the non-equilibrium
statistical relativistic diffusion model with three sources, which
include a central source and two fragmentation
sources~\cite{Wolschin:2011mz, Wolschin:2013pu}, the multisource
thermal model uses two cylindrical sources to describe the rapidity
or pseudorapidity distributions. In the present work, we embed the
exponential distribution into the geometrical picture of the
multisource thermal model to describe the transverse momentum
spectra in $pp$ collisions at RHIC and LHC energies. In the rapidity
space,  final-state particles emit from the sources, which are at
different locations due to stronger longitudinal
flow~\cite{BraunMunzinger:1994xr, BraunMunzinger:1995bp,
Feng:2011zze}. The improved multisource thermal model can reproduce
the experimental results of pions, kaons and protons. The rapidity
width is naturally taken into account by the source distribution. By
analysing systematically the corresponding pseudorapidity
distributions and fitting the experimental data, the model
parameters are fixed. The $<P_{T1}>$ is a function of
ln$\sqrt{\mathrm{\it s_{NN}}}$, other parameters are constant
values.

Summarizing up, the transverse momentum distributions of pions,
kaons and protons produced in $pp$ collisions at RHIC and LHC
energies have been studied systematically in the improved
multisource thermal model, which can reproduce $P_{T}$ spectra. Our
investigations indicate the improved model is successful in the
description of hadron productions. At the same time, it is found
that the free parameter $<P_{T1}>$ used in the calculations exhibits
linear dependences on ln$\sqrt{\mathrm{\it s_{NN}}}$. According to
the parameter change pattern, we give the model predictions of the
transverse momentum spectra of pions, kaons and protons produced at
potential future $pp$ colliders operating at $\sqrt{\mathrm{\it
s_{NN}}}$ = 33 and 100 TeV. With more accumulated data or higher
expected energy scale, the high-luminosity LHC and the
next-generation $pp$ colliders offer great opportunities for the
search for physics up to and beyond TeV scale.

\section*{Acknowledgments}

 This work is supported by National Natural
Science Foundation of China under Grants No. 11247250 and No.
11575103, Shanxi Provincial Natural Science Foundation under Grant
No. 201701D121005, and Scientific and Technological Innovation
Programs of Higher Education Institutions in Shanxi (STIP) Grant No.
201802017.

 \vskip1.0cm


\newpage

\begin{figure}[htbp]
\begin{center}
\vskip -0.cm
\includegraphics[width=0.85\textwidth]{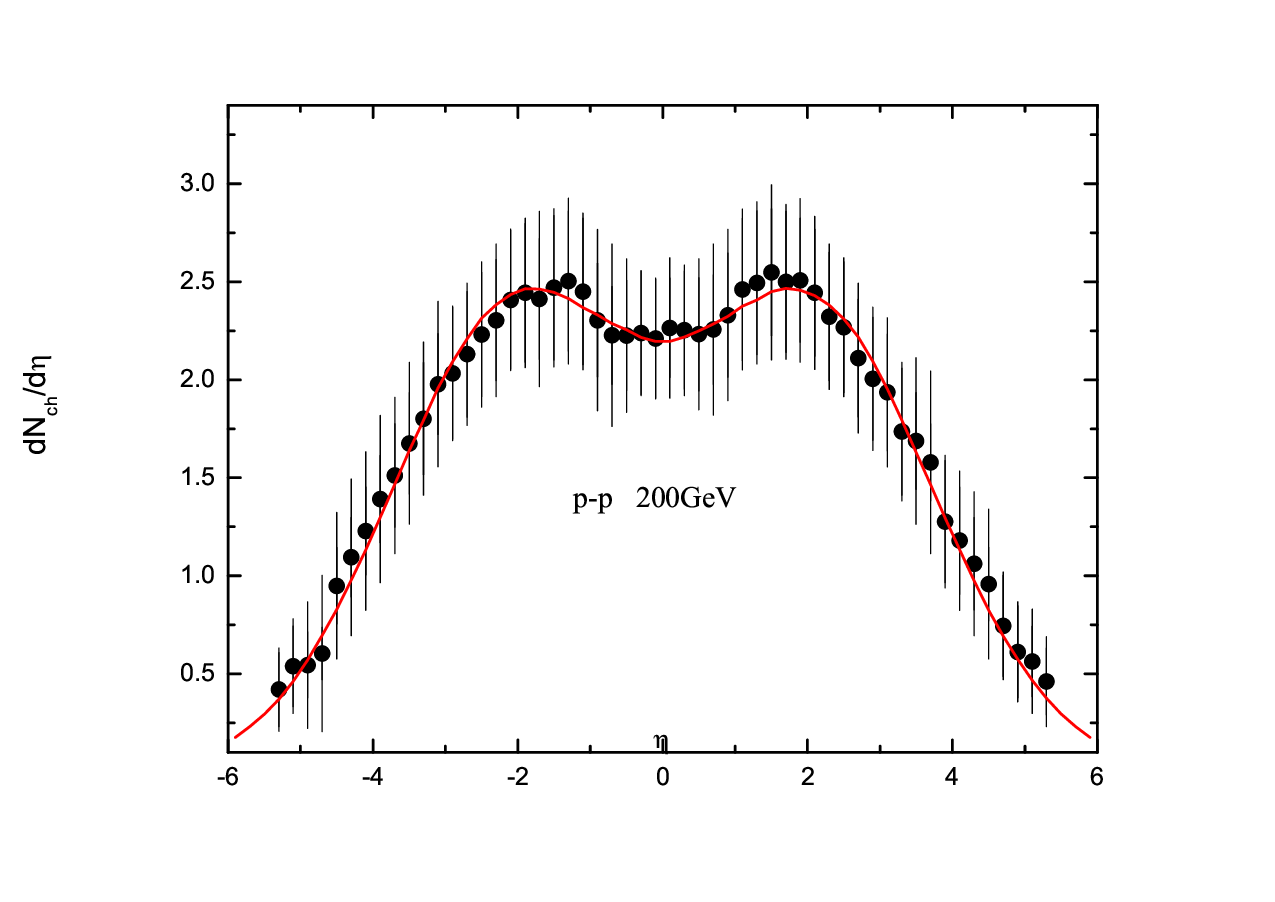}
\end{center} \vskip-0.7cm
 \caption{(Color online) Pseudorapidity distribution of charged particles produced in $pp$ collisions at
$\sqrt{\mathrm{\it s_{NN}}}$ =200 GeV.  Filled circles represent
experimental data measured by the PHOBOS
Collaboration~\cite{Alver:2010ck}. The theoretical result is
presented by a curve.}
\end{figure}

\begin{figure}[htbp]
\begin{center}
\vskip -0.cm
\includegraphics[width=0.85\textwidth]{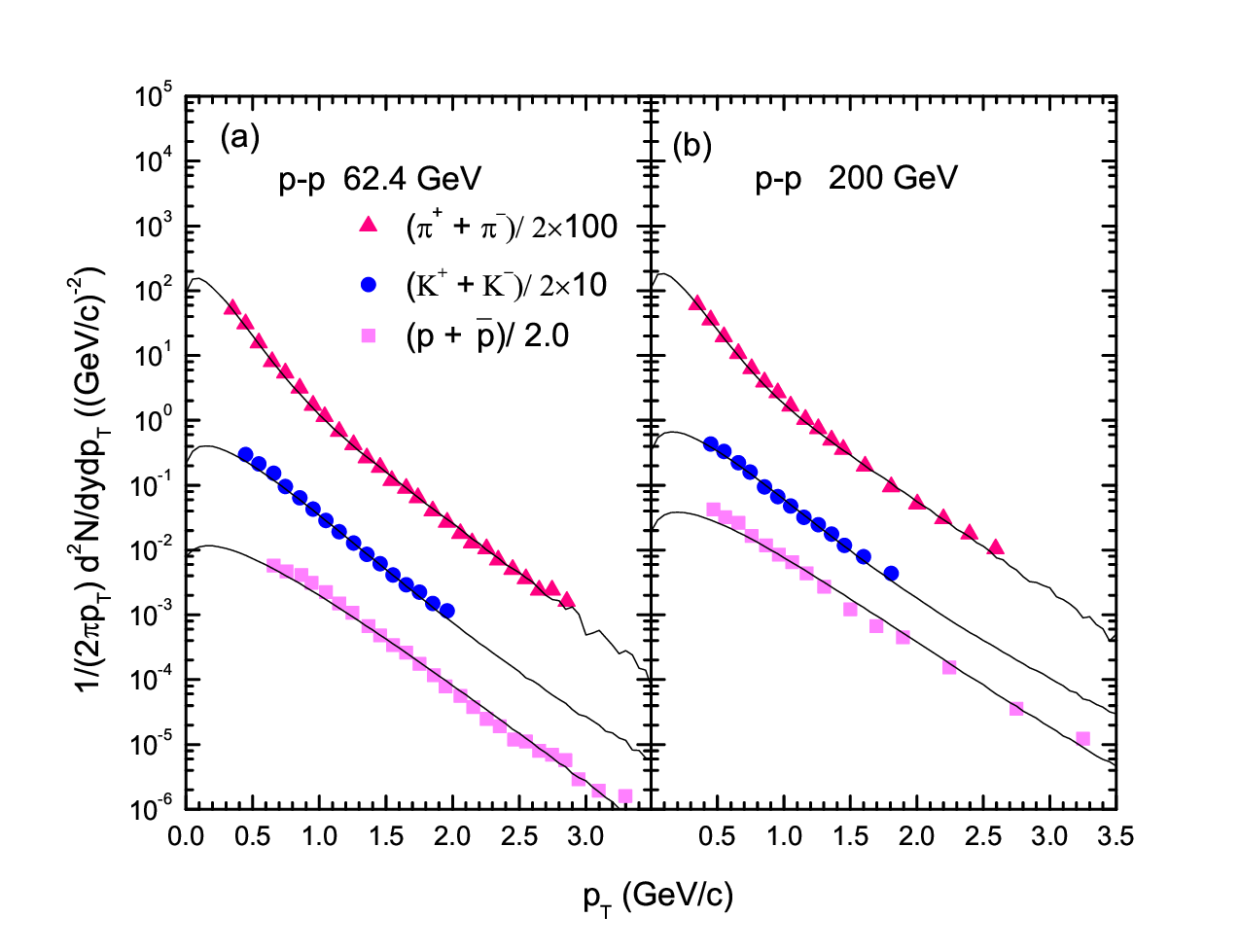}
\end{center} \vskip-0.7cm
 \caption{(Color online) Transverse momentum spectra of identified charged
hadrons (pions, kaons, protons) in $pp$ collisions at
$\sqrt{\mathrm{\it s_{NN}}}$ = 62.4 and 200 GeV.  Experimental data
from the PHENIX Collaboration~\cite{Adare:2011vy, Adare:2008qb,
Adare:2010fe} and the STAR Collaboration~\cite{Abelev:2006cs,
Adams:2006nd} are shown by the scattered symbols. The model results
are shown by the solid line.}
\end{figure}

\begin{figure}[htbp]
\begin{center}
\vskip 0.7cm
\includegraphics[width=0.85\textwidth]{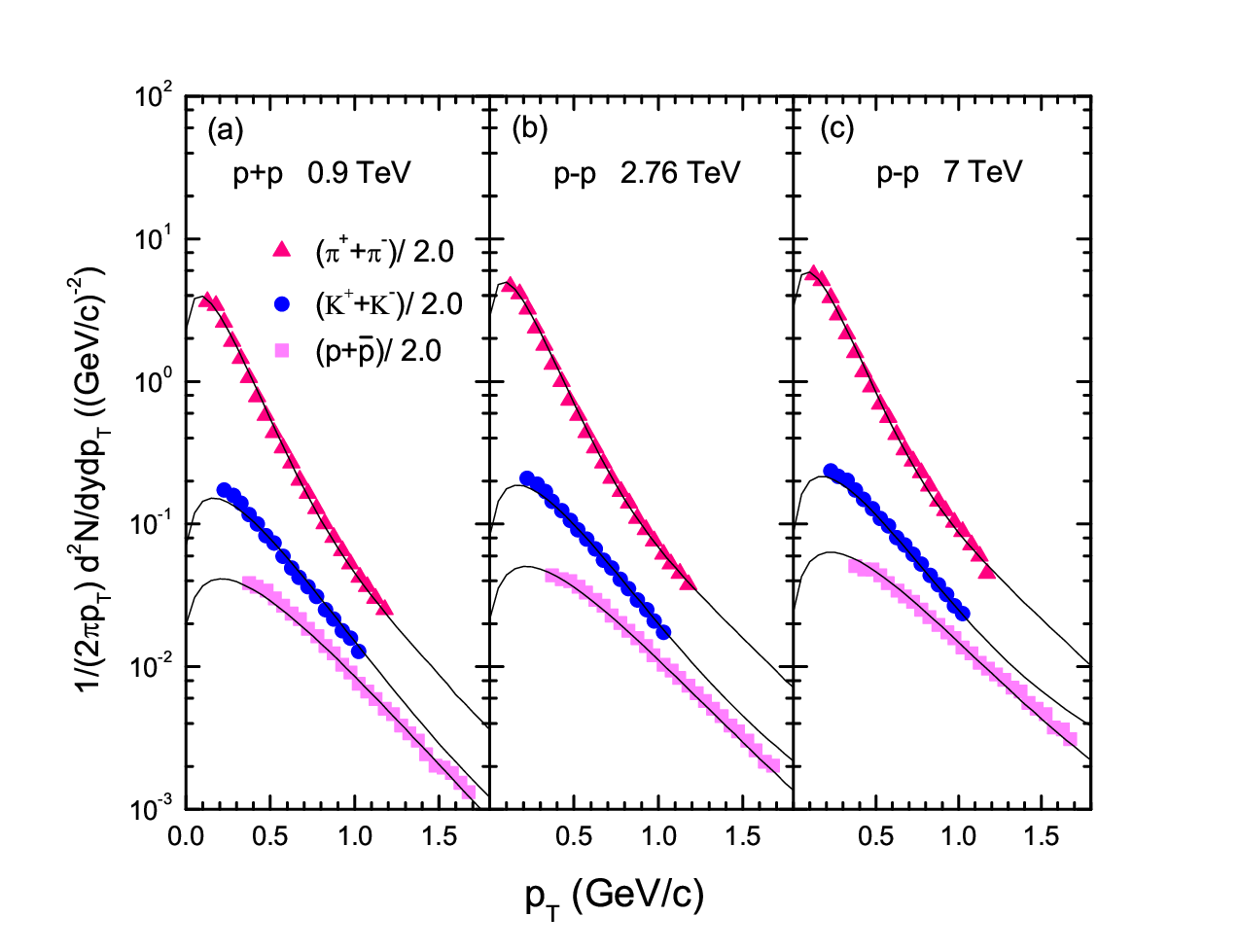}
\end{center} \vskip-1.1cm
 \caption{(Color online) Transverse momentum spectra of identified charged
hadrons (pions, kaons, protons) in $pp$ collisions in the range
$|y|<1$, at $\sqrt{\mathrm{\it s_{NN}}}$ = 0.9, 2.76 and 7 TeV.
Experimental data measured by the CMS
Collaboration~\cite{Chatrchyan:2012qb} are shown by the scattered
symbols.  The model results are shown by the solid line.}
\end{figure}

\begin{figure}[htbp]
\begin{center}
\includegraphics[width=0.85\textwidth]{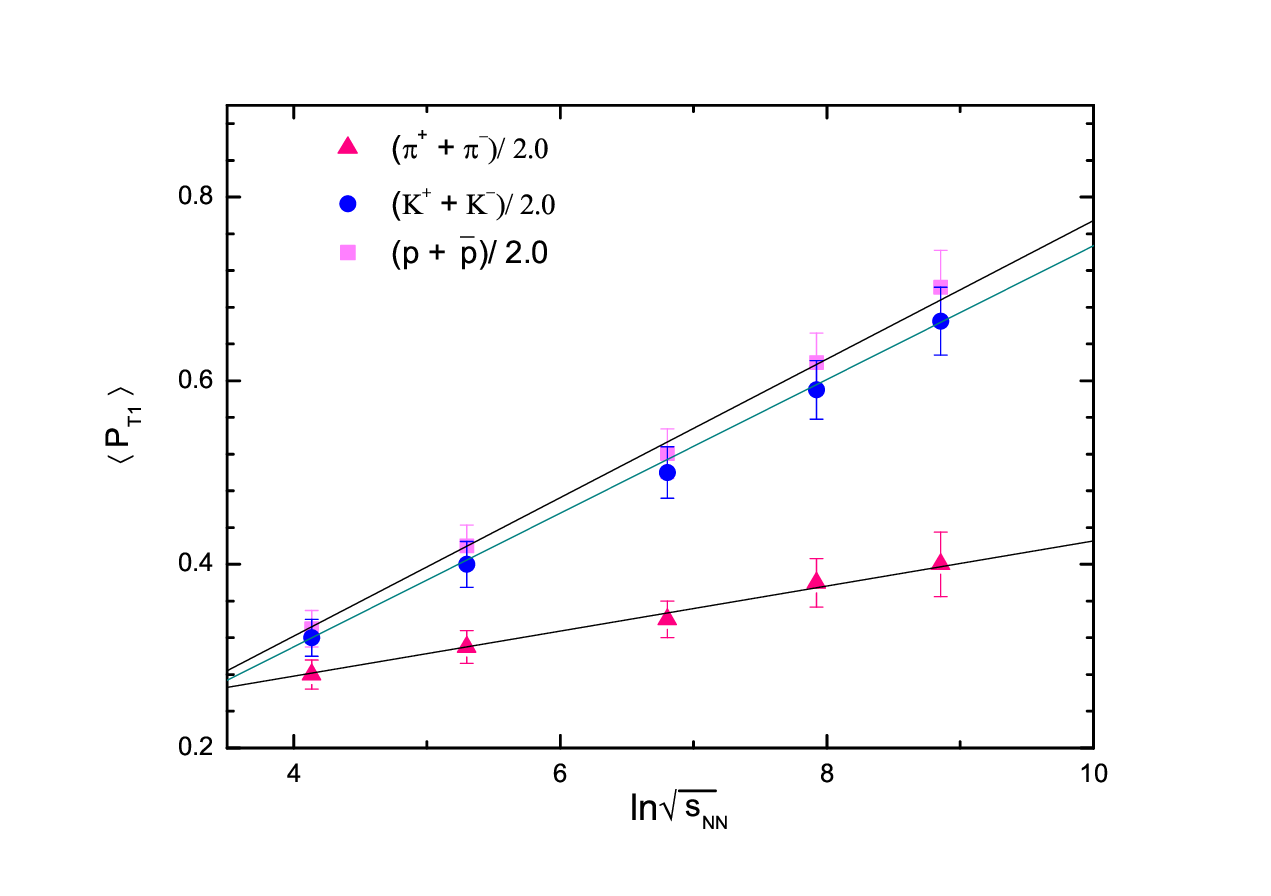}
\end{center} \vskip-1.5cm
\caption{(Color online) The dependence of the different parameters
on $\ln{\sqrt{\mathrm{\it s_{NN}}}}$. The symbols represent the
 parameter values used in the calculations for different experimental
collaborations.  The solid lines denote the fitted results. }
\end{figure}

\begin{figure}[htbp]
\begin{center}
\includegraphics[width=0.85\textwidth]{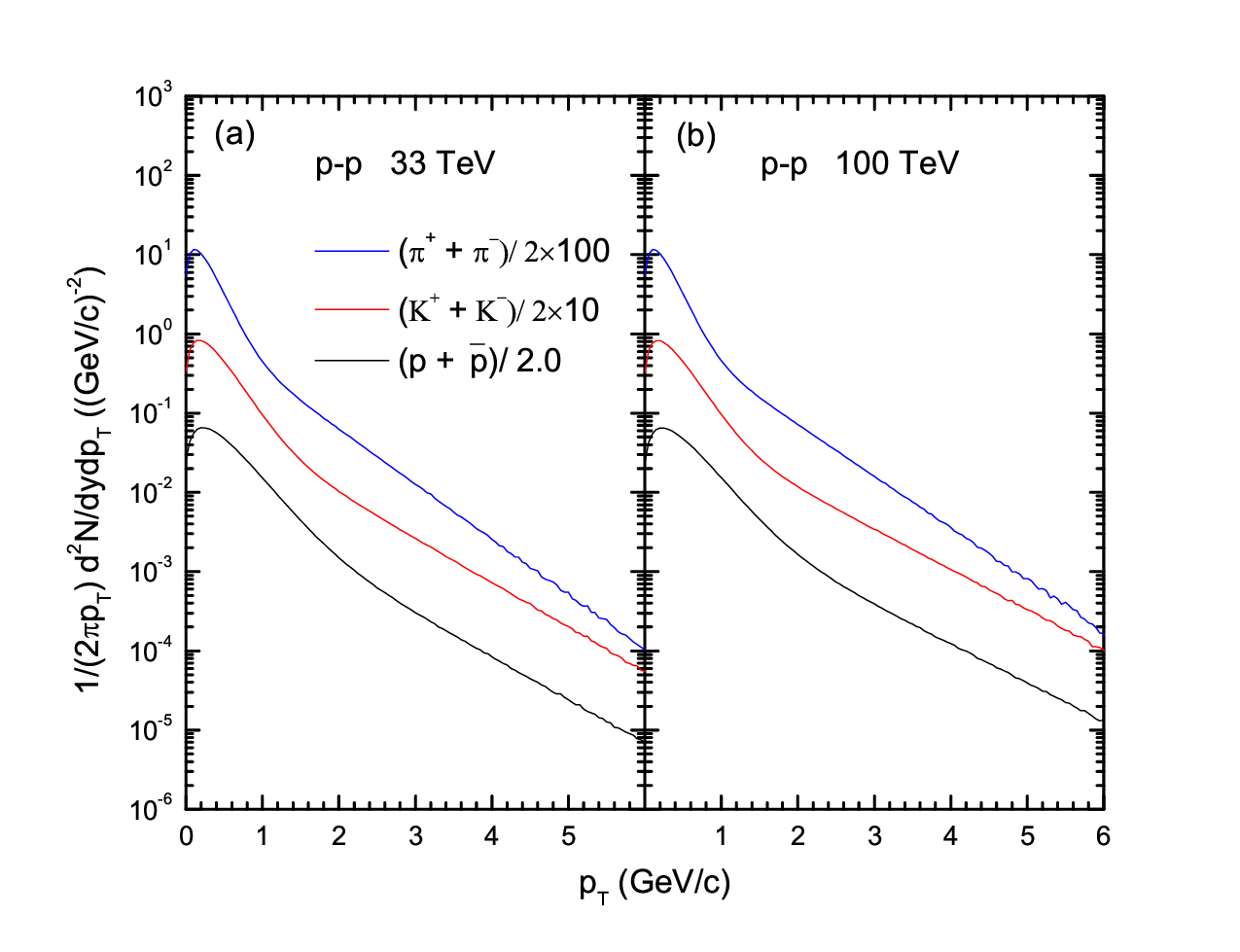}
\end{center} \vskip-1.0cm
\caption{(Color online) The the transverse momentum spectra of
pions, kaons and protons in $pp$ (or $p\overline{p}$)  collisions at
$\sqrt{\mathrm{\it s_{NN}}}$ = 33 TeV and 100 TeV.}
\end{figure}


\begin{thebibliography}{99}
\bibitem{Aad:2016mok}
  G~Aad {\it et al.} [ATLAS Collaboration]
  {\it Phys.\ Lett.\ B} {\bf 758} 67 (2016).
\bibitem{Mishra:2015pta}
  A~N~Mishra, P~Sahoo, P~Pareek, N~K~Behera, R~Sahoo and B.~K.~Nandi
  arXiv:1505.00700 [hep-ph].
\bibitem{Cleymans:2016qnc}
  J~Cleymans, B~Hippolyte, H~Oeschler, K~Redlich and N~Sharma
  arXiv:1603.09553 [hep-ph].


\bibitem{Becattini:2000jw}
  F~Becattini, J~Cleymans, A~Keranen, E~Suhonen and K~Redlich
  {\it Phys.\ Rev.\ C} {\bf 64} 024901 (2001).
\bibitem{Broniowski:2002ea}
  W~Broniowski and W~Florkowski
  hep-ph/0202059.
\bibitem{Broniowski:2002nf}
  W~Broniowski, A~Baran and W~Florkowski
  {\it Acta Phys.\ Polon.\ B} {\bf 33} 4235 (2002).



\bibitem{Abelev:2008ab}
  B~I~Abelev {\it et al.} [STAR Collaboration]
  {\it Phys.\ Rev.\ C} {\bf 79} 034909 (2009).
\bibitem{Hagedorn:1965st}
  R~Hagedorn
  {\it Nuovo Cim.\ Suppl}.\  {\bf 3} 147 (1965).
\bibitem{Abgrall:2013qoa}
  N~Abgrall {\it et al.} [NA61/SHINE Collaboration]
  {\it Eur.\ Phys.\ J.\ C} {\bf 74} 2794 (2014).
\bibitem{Liu:2002ws}
  F~H~Liu
  {\it Phys.\ Rev.\ C} {\bf 66} 047902 (2002).
\bibitem{Wolschin:2011mz}
  G~Wolschin
  {\it Europhys.\ Lett}.\  {\bf 95} 61001 (2011).
\bibitem{Alver:2010ck}
  B.~Alver {\it et al.}  [PHOBOS Collaboration],
  Phys.\ Rev.\ C {\bf 83}, 024913 (2011).
\bibitem{Adare:2011vy}
  A~Adare {\it et al.} [PHENIX Collaboration]
  {\it Phys.\ Rev.\ C} {\bf 83} 064903 (2011).
\bibitem{Adare:2008qb}
  A~Adare {\it et al.} [PHENIX Collaboration]
  {\it Phys.\ Rev.\ D} {\bf 79} 012003 (2009).
\bibitem{Adare:2010fe}
  A~Adare {\it et al.} [PHENIX Collaboration]
  {\it Phys.\ Rev.\ D} {\bf 83} 052004 (2011).
\bibitem{Abelev:2006cs}
  B~I~Abelev {\it et al.} [STAR Collaboration]
  {\it Phys.\ Rev.\ C} {\bf 75} 064901 (2007).
\bibitem{Adams:2006nd}
  J~Adams {\it et al.} [STAR Collaboration]
  {\it Phys.\ Lett.\ B} {\bf 637} 161 (2006).
\bibitem{Chatrchyan:2012qb}
  S~Chatrchyan {\it et al.} [CMS Collaboration]
  {\it Eur.\ Phys.\ J.\ C} {\bf 72} 2164 (2012).

\bibitem{Tiwari:2013wga}
  S~K~Tiwari and C~P~Singh
  {\it Adv.\ High Energy Phys}.\  {\bf 2013} 805413 (2013)

\bibitem{Huang:2003jv}
  H~Z~Huang
  {\it J.\ Phys.\ G} {\bf 30} S401 (2004).

\bibitem{Liu:2014xna}
  F~H~Liu, T~Tian, H~Zhao and B~C~Li
  {\it Eur.\ Phys.\ J.\ A} {\bf 50} 62 (2014).

\bibitem{Wang:2011zzh}
  E~Q~Wang, H~R~Wei, B~C~Li and F~H~Liu
  {\it Phys.\ Rev.\ C} {\bf 83} 034906 (2011).

\bibitem{Li2013AHEP} B~C~Li, Y~Y~Fu, L~L~Wang, F~H~Liu Adv.\ High Energy Phys.\
{\bf 2013}, 908046 (2013).

\bibitem{Wolschin:2013pu}
  G~Wolschin
  {\it J.\ Phys.\ G} {\bf 40} 045104 (2013).
\bibitem{BraunMunzinger:1994xr}
  P~Braun-Munzinger, J~Stachel, J~P~Wessels and N~Xu
  {\it Phys.\ Lett.\ B} {\bf 344} 43 (1995).
\bibitem{BraunMunzinger:1995bp}
  P~Braun-Munzinger, J~Stachel, J~P~Wessels and N~Xu
  {\it Phys.\ Lett.\ B} {\bf 365} 1 (1996).
\bibitem{Feng:2011zze}
  S~Q~Feng and Y~Zhong
  {\it Phys.\ Rev.\ C} {\bf 83} 034908 (2011).

\end{thebibliography}
\end{document}